# High performance Lunar landing simulations

Jérémy Lebreton*, Roland Brochard, Nicolas Ollagnier,
Matthieu Baudry, Adrien Hadj Salah, Grégory Jonniaux, Keyvan Kanani, Matthieu Le Goff, Aurore Masson

*Airbus Defence & Space, Toulouse, France*
* Corresponding Author, surrender.software@airbus.com

**Abstract**

Autonomous precision navigation to land onto the Moon relies on vision sensors. Computer vision algorithms are designed, trained and tested using synthetic simulations. High quality terrain models have been produced by Moon orbiters developed by several nations, with resolutions ranging from tens or hundreds of meters globally down to few meters locally. The SurRender software® is a powerful simulator able to exploit the full potential of these datasets in raytracing. New interfaces include tools to fuse multi-resolution DEMs and procedural texture generation. A global model of the Moon at 20m resolution was integrated representing several terabytes of data which SurRender can render continuously and in real-time. This simulator will be a precious asset for the development of future missions.

**Acronyms/Abbreviations**
API: Application Programming Interface
CPU: Central Processing Unit
DEM: Digital Elevation Model
DOM: Digital Ortho Model
FOV: Field Of View
GPU: Graphical Processing Unit
GSD: Ground Sample Distance
I/O: Input/Output
LOLA: Lunar Orbiter Laser Altimeter
LRO: Lunar Reconnaissance Orbiter
LSB: Least Significant Bit
NVME: Non Volatile Memory Express
PDS: Planetary Data System
PSF: Point Spread Function
RAM: Random Access Memory
SSD: Solid State Drive
SuMoL: SurRender Modelling Language
SSTMP: Solar System Treks Mosaic Pipeline
TB: Terabyte
VBN: Vision-Based Navigation

## 1. Introduction

*Airbus Defence & Space* is participating in a worldwide effort led by space agencies to take the next steps to the Moon. The new generation of spacecraft will embed vision-based navigation (VBN), which offers a reliable and efficient solution for autonomous precision landing. The development and validation of algorithms is based on simulated images generated from real terrain models. Absolute navigation requires particularly representative simulations because the navigation is performed relative to a terrain database. Therefore terrain models resolution and accuracy are performance drivers for the mission.

For this reason *Airbus Defence & Space* decided to develop a global image simulator of the Moon targeting unprecedented spatial resolution and real time. The simulator uses the high performance rendering engine SurRender software® and open source datasets from past Lunar missions.

Our first approach is to analyze lunar topographic data from NASA Lunar Reconnaissance Orbiter. LRO Narrow Angle Camera offers global coverage with multiple image pairs. This opens the possibility of stereo reconstruction using NASA Solar System Treks Mosaic Pipeline. We assess the achievable image quality targeting at least 5m resolution on selected sites. In a parallel approach we download datasets from China Chang'e-2 mission, consisting of high quality DEMs (Digital Elevation Models, or equivalently Digital Terrain Models) with native resolution of 20m and orthoimages at 7m resolution. We assemble the mosaic using SurRender highly efficient *build_conemap* algorithm which implements a pyramidal scheme and maps the spheroid using cube maps[1].

The next step is to simulate synthetic views taking profit of the unique performances of SurRender raytracer. Because it faces challenging requirements (gigantic datasets, physical realism, etc.) SurRender implementation competes with the state-of-the art of the computer graphics industry. Major optimizations are introduced for these demanding simulations, as well as new tools, interfaces and extended support of data standards. We achieve the goal of simulating full-field 1024x1024 images in raytracing at 15 Hz on a 16 cores CPU with low compromise on quality. Furthermore simulations can be augmented with synthetic features - procedural textures, rocks and craters - to artificially increase the DEM resolution.

---

[1] *Cube map*s are commonly used to implement *environment maps* in computer graphics as it solves some flaws of *spherical mapping* such as distortions and artifacts close to the poles.

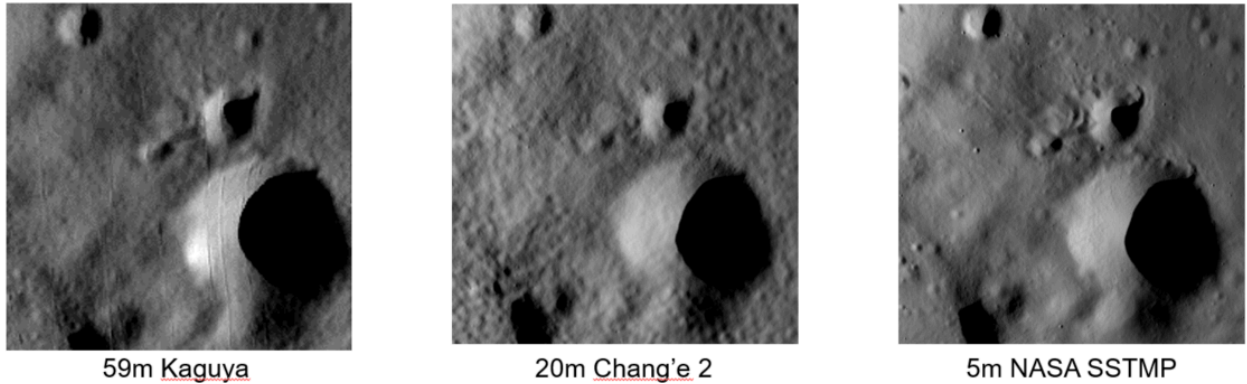

Fig. 1. Rendering of a Lunar crater under the same illumination conditions with 3 different datasets

## 2. Review of terrain datasets

Multiple terrain datasets are available in public repositories from international space agencies based on the Planetary Data Systems (PDS) standard. To design a realistic simulator, both image products (albedo/reflectance maps or ortho-images) and Digital Elevation Models are needed. The Ground Sample Distance (GSD, e.g. resolution) shall be as fine as possible and the coverage should ideally be global. A good image quality is desirable and in particular datasets should not have any artifacts.

We browsed data sources including European SMART-1, JAXA Kaguya (SELENE), ISRO Chandrayaan-2, Nasa LRO and CNSA Chang'e-2. The primary data product from these missions is orthoimages. High resolution DEM are computed from stereo reconstruction using various algorithms (see for instance review by [2]). A rule of thumb is that the resolution of DEMs is at best 4 times less than the native image resolution. We focus on the LRO, Kaguya and Chang'e-2 datasets. LRO/LOLA is an altimeter. Kaguya and LROC datasets have been coregistered with LOLA.

Table 1. Lunar DEM datasets

|  | Coverage | Resolution | Ref |
|---|---|---|---|
| LOLA | Global | 118m | [4] |
| Kaguya(+LOLA) | [60°S,+60°N] | 59m | [5] |
| Chang'e-2 | Global | 20m | [6] |
| Chandrayaan-2 | Local tiles | 10-25m | [7] |
| LROC(+LOLA) | Local tiles | 2-5m | [8] |

High resolution DEMs can be computed from 1m or less LROC images with NASA SSTMP pipeline [1]. In particular, datasets are available for selected sites on the LRO website. However stereo reconstruction is a difficult task and there is no off-the-shelf tool capable of rendering high quality DEM globally. Quality of 3D reconstruction is impacted by the algorithm performance and also by the image acquisition mode. For instance we find that the stereo accommodation of the Chang'e-2 yields robustness to illumination conditions because image pairs are taken almost simultaneously. Quality of reconstruction, accuracy and georeferencing are hard to evaluate because of lack of ground truth and metrics. In Figure 1, we show the same crater simulated with datasets from three different data sources. The 59m Kaguya+LOLA image exhibits spurious lines characteristics of stereo reconstruction artifacts. The Chang'e-2 image includes high resolution textures which are not apparent in the 5m LRO image stereo-processed with NASA SSTMP and could be artifacts. There are no obvious metrics to distinguish which is closer to the ground truth.

The selected input dataset for the simulator is the 20 m Chang'e-2 DEM and 118 m albedo map from Kaguya/SELENE. High resolution Chang'e-2 or LRO orthoimages are not exploitable directly because they have not been corrected for shading. To achieve higher resolution the DEM is merged with 5 m resolution LRO tiles locally.

## 3. SurRender software

The functionalities of the SurRender software have been presented in two previous papers [9, 10]. We introduce here new features which were introduced to enable the creation of a global model of the Moon in high resolution for real time applications.

*3.1 SuMoL textures*
A new feature of SurRender is SuMoL textures. Rather than being limited to image files, textures can be defined procedurally, e.g. using mathematical functions or code described in a script. For example one possible application is the fractal generation of textures or elevation models at all scales.

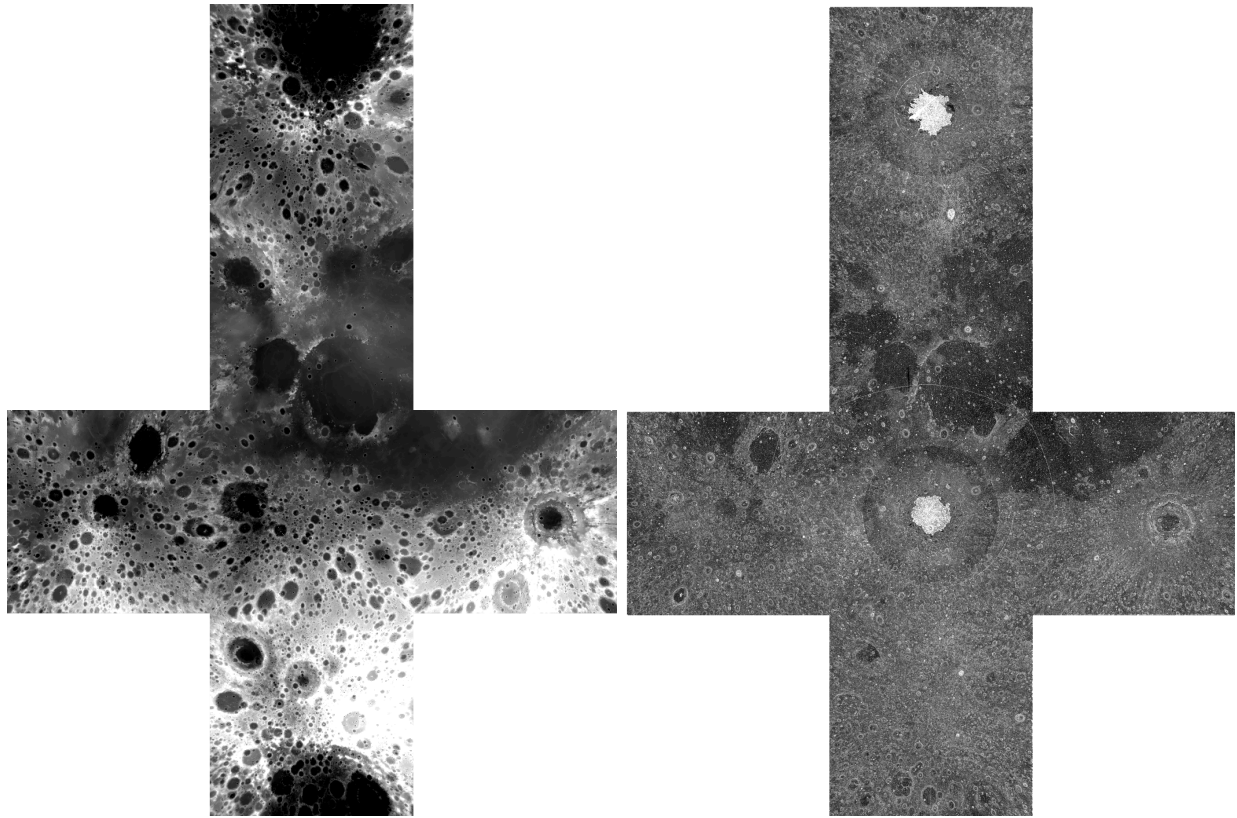

Fig. 2. Cube maps of the Moon elevation map (left) and cone map (right) computed from the Chang'e-2 20m dataset. This representation enables storing the entire Moon at arbitrary resolution (here 20m) with minimal distortions caused by cartographic projection and it is well suited for raytracing.

Another important application is that SuMoL textures enable the fusion of DEMs of multiple resolutions.
Procedural textures can also be used to define texture transformations at runtime with minimal cost. SurRender now supports cube mapping textures (cubemaps) Cube mapping involves less geometric distortions, faster texture lookup and ray intersections computing, compared to spherical or cylindrical projections. With this interface, there is no singularity at the pole which opens the possibility to render the pole without any projection artifacts.

*3.2 Interfaces*
SurRender interface with PDS textures is implemented in the tools *build_conemap* (for planar geometry), *build_conemap_spherical* (for spheroids) and *build_conemap_cubemap*. They generate acceleration structures called *conemaps* (representing local tangents) and *heightmaps* (relative elevation) which are encoded using a pyramidal scheme in format .BIG. With SurRender 8, the functionalities have been upgraded, in particular:
- Extended PDS support, GeoTIFF, geomapping
- New projection models (conic, equirectangular, mercator, …) described as SuMoL textures
- PDS support in SurRender server (for albedo maps)
- Fusion of DEM.

The latter function is very important. Using SuMoL textures, it is now possible to merge datasets of different resolutions and projection models. The overlap or junctions between tiles are controlled using SuMoL scripts, lookup tables and masks. The software *big_viewer* is a graphical user interface that displays .BIG files very fast thanks to the memory mapping implementation of SurRender. For instance in Figure 2, the global DEM model is displayed and it can be zoomed with no latency down to the native resolution (and beyond). It works on network drives too which is very convenient for models of several terabytes.

*3.3 Simulation input preparation*
A global DEM is assembled. We merged Chang'e-2 datasets, which consist in 188 (DEM) + 750 (DOM) tiles that mosaic the sphere with several different projection systems. From the poles to the equator stereographic projection, conic projection and Mercator projection are respectively used in the input datasets.

The input dataset weighs about 500 GB. The cubemaps are optimized for simulation, e.g. the main objective is not to save storage space, but to be IO-optimized (less I/O required) and computation-optimized (less computation required during simulation). The reconstruction and optimization process is both computationally and I/O intensive. This process was performed on a dedicated workstation with powerful processing power and efficient storage. Each step is heavily multi-threaded. The bottleneck for merging tiles is essentially the I/O and this step takes about 4 hours. The conemap computation is limited by the CPU and takes about 18 hours. The entire process requires about 24 hours. After optimization the reconstructed global DEM requires about 1.5 TB of disk space. The DOM (Digital Ortho Model) also takes 1.5 TB. File sizes are 5 TB and 2.5 TB respectively because they are sparse.

Last, fusion with a high resolution LRO DEM is performed procedurally in order to match the camera resolution at low altitudes.

## 3. Results

### 3.1 Computing performance

The dataset (DEM+DOM) represents 3 TB stored on a SSD drive. Simulations are performed at ~15 Ghz with a dedicated workstation with a 16 core-CPU (AMD Ryzen 9 5950X) and plenty of NVME SSD storage. The dataset is so large that it is impractical to use the OpenGL pipeline which exploits the GPU cache memory. Simulations are made in raytracing targeting either high quality or real time. High quality means that for a large enough number of rays per pixel, there is no numerical noise owing to the random sampling of pixels in the images. SurRender implements a highly efficient ray sampling method by focusing on the signal that really impacts image quality, thus good quality images are achieved even with a limited number of rays. In real time with 1 ray per pixel and no PSF images are rendered at 15Hz (low quality for real-time tests). Computational time scales approximately with the number of rays, although rendering speed depends on scene content and geometry. A high quality image with 100 rays per pixel is rendered in 5 sec. An example of rendering is shown in Figure 3 together with the difference between a high quality image and a medium quality image (10 rays/pixel). The difference is not visible by the naked eye but residuals between the two images highlight minor differences, especially along sharp edges. The dynamics of this image is 8 bits e.g. pixel values range from 0 to 255 LSB. The residual between the two images has a standard deviation of only 2.3 LSBs after normalizing the images by the mean and excluding pixels which show no difference.

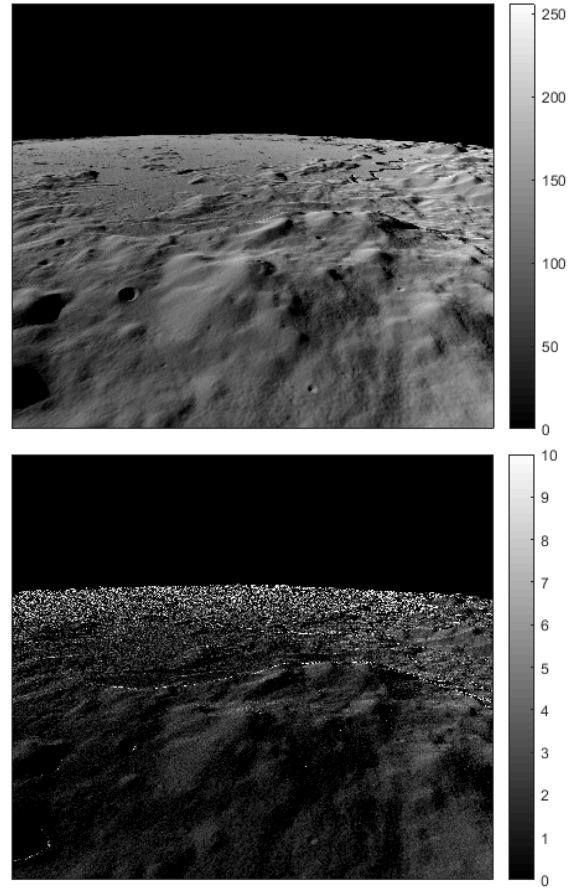

Fig. 3. Impact of rendering quality vs computing performance. Top: high quality image (100 rays/pixel). Bottom: residuals from the difference between high quality and medium quality image (10 rays/pixel)

### 3.2 Simulation highlights

Multiple landing scenarios were simulated using SurRender with this dataset. The camera is a wide FoV pinhole camera with a PSF model and a 1024x1024 detector. More detailed camera models are available but were not needed. For example we have retrieved and processed metadata from Chang'e-3 landing and simulated views acquired with the landing camera. An extract from the sequence is shown in the first panel of Figure 4 with the 20 meter DEM only. The landscapes and surface features reproduce very well the real images. In the middle panel is an example extracted from a touch and go video sequence simulated from 1500 km down to 20 km when the GSD approaches the pixel resolution of the 1024x1024 sensor. The complete video is available on the SurRender website *https://www.airbus.com/en/products-services/space/customer-services/surrendersoftware*. In order to cope with insufficient resolution of the 20 m DEM below a few kilometers of altitude, a high resolution LRO tile at 5 m resolution can be merged with Chang'e-2. This is illustrated in the bottom panel of Figure 4. The merging

IAC-22,A3,IP,44,x71795                                                                                 Page 4 of 8

is procedurally defined with SuMoL textures. A few artifacts are evident in the images, they are intrinsic to the datasets.

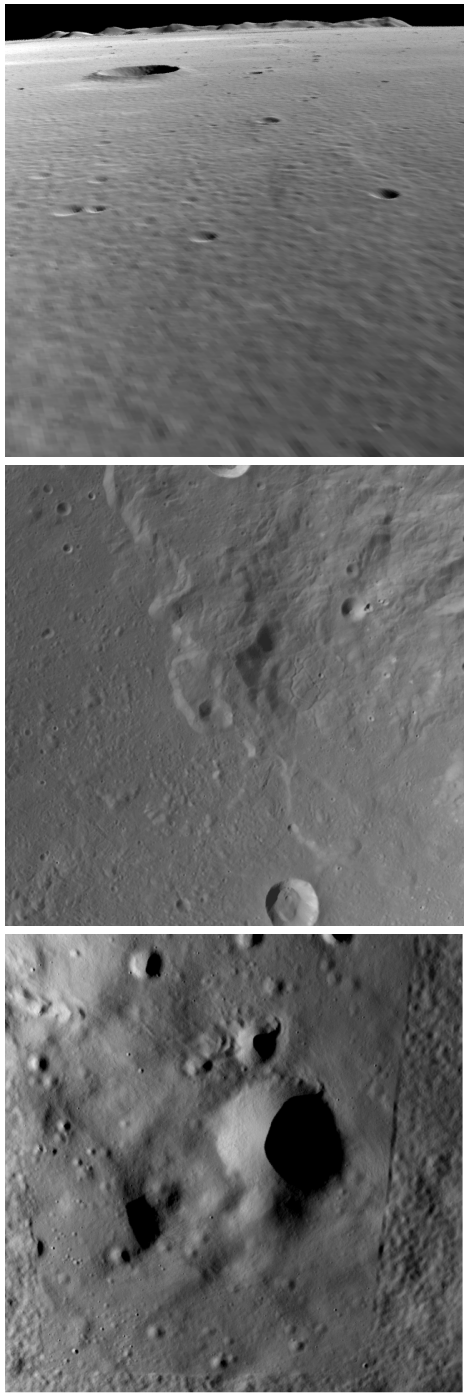

Fig. 4. A sample of images simulated with the Chang'e 2 20m DEM.
Top: extract from Chang'e-3 descent trajectory imaged with the landing camera. Middle: random view of the Moon extracted from the touch and go video presented in the accompanying interactive presentation. Bottom: the DEM is fused with a high resolution LRO tile.

*3.3 Illumination, contrast & Hapke backscattering*
Image radiometry is very much impacted by the BRDF and Sun illumination. A critical phenomenon is the opposition effect that arises when the observed surface is illuminated from directly behind the camera, or in other words when the phase angle approaches zero. In this case there is a sudden increase in brightness and simultaneously a sudden drop in contrast in the image: all the lines-of-sights become equally reflective.
The Hapke model [11] is the BRDF largely accepted for the Moon because it captures well the reflectance properties of a surface covered with many dust particles. It is particularly prone to this effect. It is worth noting that the effect extends beyond near-zero phase angles. For instance in Figure 4, we show a rendering of the same scene illuminated either from the back or from the upper direction.

Analyzing the illumination conditions at the poles is important in the perspective of future missions. This can be done using satellite imagery directly [12] or by simulation. For example [13] used LOLA datasets to sample the South pole illumination over a year. SurRender simulates precisely the geometry and radiometry of shadows including soft shadows under the effect of BRDF and the Sun angular size. The elevation model is among the most accurate available. For all of these reasons, the simulator can be used to reliably analyze the illumination conditions on a Lunar landscape. Figure 5 shows the illumination at the Lunar South pole integrated over time by sampling the Sun elevation and direction over a year. Although there are permanently illuminated zones at the South pole, the number of illumination hours is limited due to shadow forecast by sometimes distant mountains because the Solar elevation is very low.

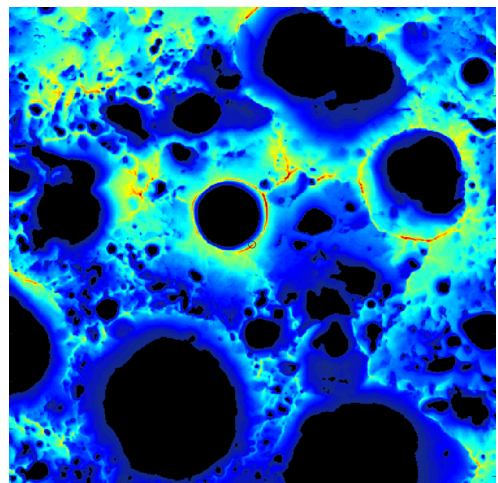

Fig. 5. Illumination at the Lunar South Pole integrated over one year

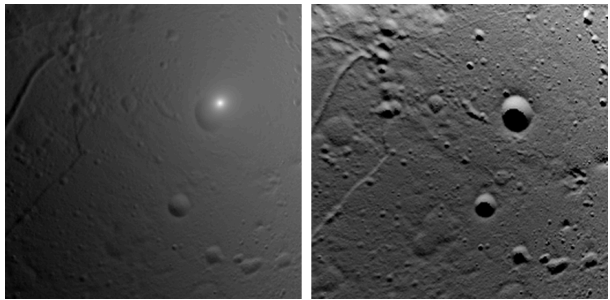

Fig. 5. Illustration of Hapke backscattering

**4. Procedural details**
For image processing validation, it can be useful to add synthetic details to an existing 3D model or DEM. A new API has been developed for SurRender. It consists of several functionalities. Illustrations are provided in Figure 6.

*4.1 Procedural DEMs and craters*
In order to add high frequency details to a terrain (or create a terrain model from scratch), one must alter a DEM (or a plane model). This is achieved by modifying the DEM locally or by using a procedural model to create textures. An analytical model can be input to generate a texture image (for albedo maps) or a PDS file (for DEM). For example Perlin noise is a procedural texture primitive often used in computer graphics to add details to objects. The API also makes it possible to add models of craters and their distribution. An example is shown in Figure 7. In this case the crater model represents circular holes on the terrain. The texture model is a Perlin noise. With this tool it becomes possible to add artificial details over many orders of magnitude.

*4.2 Distribution of meshes (boulders)*
For various applications, it is useful to load numerous 3D models into the scene. This can be tricky computationally because each 3D mesh occupies cache memory so a lot of them quickly saturates RAM. The new API enables instantiating many times the same meshes with different scale, attitudes and positions. Mesh instances are agglomerated in a single compact object. We tested that with this method, SurRender can load as many as 1,000,000 boulders in a scene on a desktop computer. In order to place the meshes at the Lunar surface, a dedicated function intersects the meshes with a DEM (or a mesh) along a specified vector, here pointing to the Moon center. Lastly, masks are used to distribute the boulders statistically, such that some zones are filled with obstacles and some are void. In Figure 6 we show an example of rendering including 20000 identical boulders randomly sized and oriented on a Moon tile.

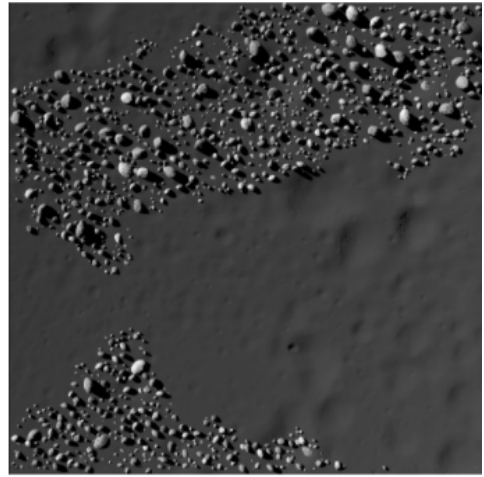
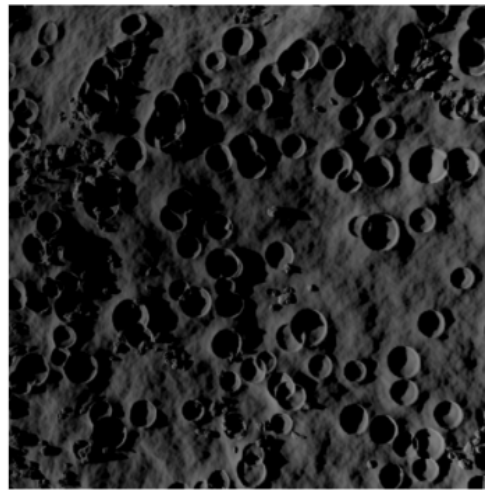
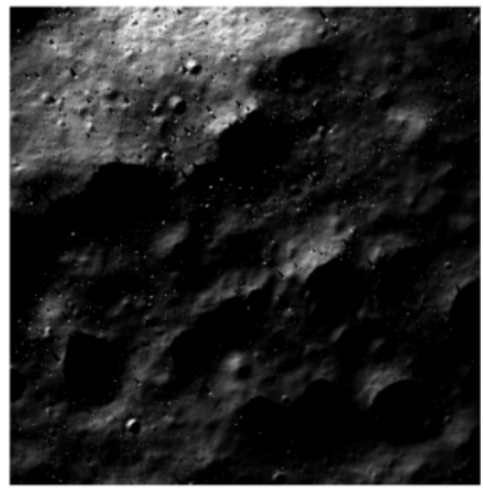

Fig. 6. Procedural texture generator.
Top: distribution of boulders and rocks. Middle: craters on a synthetic DEM. Bottom: synthetic albedo and boulders on a LRO tile at an altitude of 500 meters.

## 6. Conclusions

The maturation of precision landing or hazard detection technologies for future Moon missions relies on simulated datasets. High quality terrain models are being released by the planetology community and new tools are needed to take advantage of these massive datasets to produce meaningful synthetic images.

We have developed a high performance Lunar landing simulator able to interface with PDS data, perform merging of multiresolution datasets and simulate physically realistic camera images using the SurRender raytracing engine. The Lunar simulator is now used in a real time closed-loop GNC simulation environment to validate the navigation chain of future missions. The images are rendered physically at subpixel level making the images adequate even for testing 3D reconstruction algorithms.

This high performance simulator will be one of the backbones of *Airbus Defence & Space* ambition to contribute to a new era of Lunar exploration. The tools are released with the new public version of SurRender 8 at the end of 2022.

The identified limitation of the simulator lies in the input datasets. The achievable resolution of satellites is not exploited to its full potential on a global scale. High resolution (few meters) LRO DEMs are only available locally. Locally datasets suffer from few artifacts caused by stereo reconstruction flaws, spurious radiometric equalization, interpolations, edge effects, etc. New algorithms are under development for example [14] are testing new deep learning-based techniques (MADNet) to reconstruct 3D models from single view imagery. A hybrid algorithm using deep learning and shape-from-shading is also presented by [15]. New automated processing pipelines are also under development to scale up the processing of LRO NAC with the NASA Ames Stereo pipeline [16]. From a conceptual standpoint, there is a lack of metrics to quantify the validity of datasets which should be further investigated. Nevertheless it shall be recognized that the work put together by agencies and the scientific community to build global Lunar cartographic products is impressive and that the quality of existing datasets is already of very high level.

## Acknowledgement

The authors thank the PILOT and the EL3 projects (ESA) for providing the application case which motivated the production and analysis of the global Lunar datasets. This paper was presented for the 73rd International Astronautical Congress (IAC), Paris, France, 18-22 November 2022. Contribution reference: IAC-22,A3,IP,44,x71795.

N.B.: This work presents the status of the SurRender Moon renderer in September 2022. Extensive developments have occurred since then, they will be presented in a future paper.